\documentstyle[twocolumn,aps,epsfig,psfig]{revtex}
\begin{document} 
\draft
\title{Initial-amplitude dependence in weakly damped oscillators}
\author{Avinash Singh}
\address{Department of Physics, Indian Institute of Technology, Kanpur-208016, India}
\maketitle
\begin{abstract} 
A pedagogically instructive experimental procedure is suggested 
for distinguishing between different damping terms in a weakly damped oscillator,
which highlights the connection between 
non-linear damping and initial-amplitude dependence.
The most common damping terms such as contact friction, 
air resistance, viscous drag, and electromagnetic damping 
have velocity dependences of the form constant, $v$, or $v^2$.
The corresponding energy dependences of the form $\sqrt{E}$, $E$, or $E\sqrt{E}$
in the energy loss equation
give rise to characteristic dependence of the amplitude decay slope  
on the initial amplitude. 
\end{abstract}
\section{Introduction}
The most commonly studied source of damping in oscillating systems
is a resistive force ${\bf F}_{\rm res} = -m {\bf v}/\tau$ proportional to velocity,
such as due to fluid drag at very low relative speed,
when the Reynolds number $R$ is of the order 1 or less.
The equation of motion involves a {\em linear} differential equation
of the type
\begin{equation}
\ddot{x} + \frac{\dot{x}}{\tau} + \omega_0 ^2 x = 0
\end{equation}
which yields an exponential decay $x_0(t)/x_0(0) = e^{-t/2\tau}$ 
for the normalized amplitude of oscillations. 
The decay is independent of the initial amplitude,
a characteristic of the linear nature of the system.
Here $x(t)$ refers to the instantaneous displacement of a block in a 
spring-mass oscillator
or the angular displacement of a physical pendulum. 

For weakly damped oscillations, 
it is convenient to consider the oscillator energy $E$ averaged over one cycle,
during which period the amplitude $x_0(t)$ is nearly constant.
With $x(t)=x_0 \sin \omega t$, we have $E = m \omega_0 ^2 x_0 ^2 /2$,
and the change in the oscillator energy over one cycle is 
$\Delta E = \int {\bf F}_{\rm res} . {\bf v} dt = - E \: T/\tau$.
This yields the linear energy loss equation 
\begin{equation}
\frac{dE}{dt} = - \frac{E}{\tau} 
\end{equation}
over long time scales compared to the oscillation time period.
The normalized energy $E(t)/E(0)= e^{-t/\tau}$ also decays exponentially, 
independently of the initial energy.

In most laboratory oscillators, however, 
non-linear terms actually dominate in the energy loss equation.
A $\sqrt{E}$ term results from the nearly velocity-independent dry friction,
for instance at the pivot of a physical pendulum or due to a brake pad, 
whereas air resistance leads to a $E \sqrt{E}$ term
as the Reynolds number is typically of the order of 1000,
in which regime the air resistance is proportional to $v^2$.
Another potential contribution to the energy loss equation  
arises from a centripetal correction to the 
normal reaction at the pivot, and therefore to the friction.
As the average centripetal force $F_c \propto \omega_{\rm max} ^2 \propto E$, 
this leads to a $E \sqrt{E}$ correction to the 
contact friction contribution in the energy loss equation.

In certain situations, 
even electromagnetic damping  in an oscillating system
leads to a $\sqrt{E}$ damping term,
as recently reported for a simple experimental setup involving a magnet oscillating 
through a coil.\cite{faraday}
Similar damping behavior is expected in 
an oscillating pendulum consisting of a copper or aluminum disk, 
periodically passing between the pole pieces of two strong magnets 
placed near the mean position, 
provided the conducting disk passes through the magnetic field 
in a very short time compared to the oscillation time period. 
On the other hand, 
a continuously acting electromagnetic damping force,
as on a conducting spinning wheel placed between two strong magnets, 
leads to a $E$ damping term. 

In practice, 
the oscillator energy loss is therefore typically described by 
the following non-linear equation 
\begin{equation}
\frac{d E}{dt} = -k \sqrt{E} - k' E - k'' E \sqrt{E} \; .
\end{equation}
In this article we describe a quantitative study of the
expected initial-condition dependence of the amplitude decay due to the
non-linear damping, and discuss a practical application of this 
initial-condition dependence.
Even though the different damping terms have characteristic decay signatures 
--- a linear decay of amplitude with time for $\sqrt{E}$ damping, 
an exponential decay for $E$ damping,
and an inverse power decay for $E\sqrt{E}$ damping --- 
a practical difficulty often encountered is how to 
distinguish between several weak damping terms present simultaneously.
Towards this end, 
we suggest a sensitive  experimental procedure for  
quantitatively identifying the separate damping contributions in Eq. (3).
Furthermore, as this procedure requires only the 
{\em initial} amplitude decay and not the full decay over long time,
it is especially useful in situations where very small oscillation amplitudes 
cannot be reliably obtained.  

Either separately, or in combination, the different damping terms 
have been considered in several earlier studies.
These include a harmonic oscillator with 
sliding friction\cite{lapidus,barratt,peters}
and viscous force,\cite{ricchiuto}
an oscillating sphere with fluid drag,\cite{gupta}
and a physical pendulum 
with air resistance,\cite{miller,crawford,nelson,squire}
dry friction,\cite{squire}
and electromagnetic damping.\cite{pederson,squire} 

\begin{figure}
\hspace*{11mm}
\psfig{file=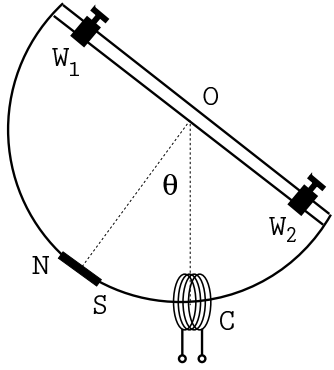,width=50mm,angle=0}
\caption{A magnet (NS) attached to an oscillating system 
passes through a coil (C) periodically, generating a series of 
emf pulses.}
\end{figure}

\section{Experimental Setup}
We have used a common laboratory setup for studying electromagnetic induction
to monitor the oscillation amplitude.
An oscillating magnet passes through a coil periodically, 
generating a series of electromotive force (emf) pulses.
In a practical realization of this concept,
a rigid semi-circular frame of aluminum, 
pivoted at the center (O) of the circle (see Fig. 1) and free to 
oscillate in its own plane about a horizontal axis through O,
has a rectangular bar magnet mounted at the center of the arc 
passing through a coil C of suitable area of cross section.
A convenient way of monitoring the induced emf pulses,
and therefore the oscillation amplitude, 
is through a PC interface, 
which can be readily realized by low-cost, convenient data-acquisition 
modules available in the market. 
The amplitude can also be directly monitored by connecting the oscillator shaft
to a precision potentiometer, and recording the instantaneous voltages through a
PC interface.\cite{squire}

The underlying electromagnetic induction phenomenon in this oscillating system 
has been discussed earlier.\cite{faraday}
The induced emf is significant only in a very narrow angular range
about the mean position,
and if the angular amplitude is not too small ($\theta_0 \gg 5^\circ$),
the angular velocity of the bar magnet 
is very nearly constant in this narrow angular range. 
The peak emf ${\cal E}_{\rm max}$ is then approximately given by
\begin{equation}
{\cal E}_{\rm max} \approx 
\left ( \frac{d\Phi}{d\theta} \right )_{\rm max}
\omega_{\rm max} \approx 
\left ( \frac{d\Phi}{d\theta} \right )_{\rm max}
\frac{2\pi}{T} \theta_0 \; ,
\end{equation}
where $T$ is the time period of (small) oscillations,
and the maximum angular velocity 
$\omega_{\rm max}=(4\pi/T) \sin (\theta_0/2) 
\approx (2\pi/T) \theta_0$,
if $\theta_0/2$ (in radians) is small compared to 1.
Thus, the peak emf ${\cal E}_{\rm max}$ provides a measure of 
the angular amplitude $\theta_0$, 
and the oscillator energy $E=\frac{1}{2} I \omega_{\rm max}^2 $
is directly proportional to $\theta_0 ^2$.

\section{Decay of Oscillator Energy and Amplitude}
While a solution of Eq. (3) is easily obtained, 
it is more instructive to proceed in steps, 
and we consider three separate cases.

\subsection{$k'=k''=0$}
When only a $\sqrt{E}$ damping term is present,
as due to contact friction and intermittent electromagnetic damping,
the solution of Eq. (3) yields a linear decay of 
the normalized oscillator amplitude with time
\begin{equation}
\frac{\theta_0 (t)}{\theta_0(0)} =
\frac{\sqrt{E(t)}}{\sqrt{E(0)}} = 1 - a t \; ,
\end{equation}
where the decay slope $a = k/2\sqrt{E(0)}$
is inversely proportional to the initial amplitude.
A phase-space plot between the normalized angular momentum
$L/L_0$ and angular displacement $\theta/\theta_0$ 
is a linear spiral,
and the number of cycles executed 
before the oscillations come to a stop 
is proportional to the initial amplitude.

\subsection{$k'' =0$} 
When a resistive force proportional to velocity is included,
which would be appropriate for low relative velocities when the Reynolds
number $R \sim 1$, the solution is modified to 
\begin{equation}
\frac{k+ k'\sqrt{E(t)} }
{k+ k'\sqrt{E(0)} } 
= e^{-k't/2} \; ,
\end{equation}
which leads to an exponential decay 
$E(t)/E(0) = e^{-k't} $ in the limit $k \rightarrow 0$.
It is instructive to consider the $k' \rightarrow 0$ limit.
Expanding the exponential term, we obtain 
\begin{equation}
\frac{\theta_0 (t)}{\theta_0(0)} =
\frac{\sqrt{E(t)}}{\sqrt{E(0)}} = 1 - a t + bt^2 + \cdots
\end{equation}
where $a = k/2\sqrt{E(0)} + k'/2$ and $b/a= k'/4$.
The $E$ damping term contributes to the deviation from linearity, 
and as $k' \rightarrow 0$, all second- and higher-order terms vanish,
leaving only the linear term, as in Eq. (5). 

\subsection{General case} 
We again consider the small-time behaviour of the 
oscillator energy $E(t)$ or amplitude $\theta_0 (t)$ 
given by  Eq. (7), and substituting in Eq. (3) we obtain 

\begin{figure}
\vspace*{-70mm}
\hspace*{-38mm}
\psfig{file=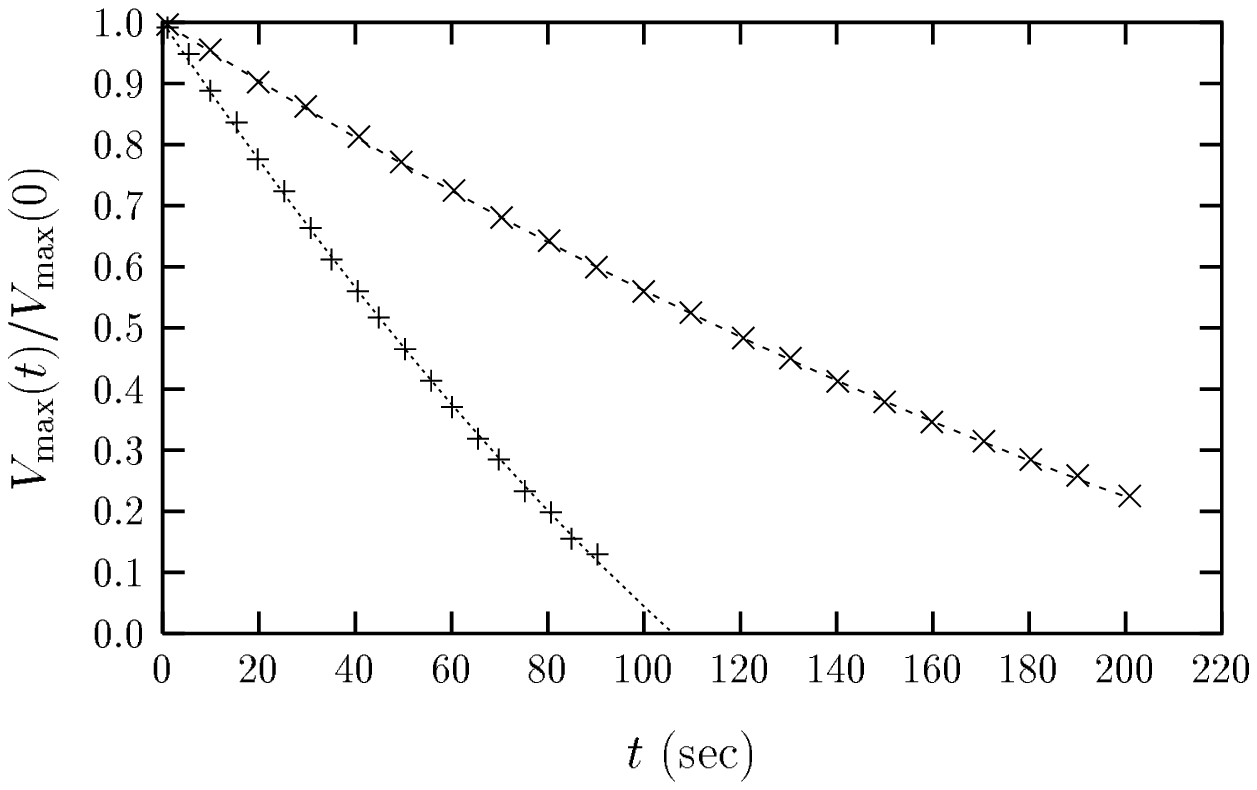,width=135mm,angle=0}
\vspace{-70mm}
\caption{
The normalized peak voltage $V_{\rm max}(t)/V_{\rm max}(0)$ vs. time $t$ 
for the short-circuit ($+$) and open-circuit ($\times$) cases.}
\end{figure}

\begin{eqnarray}
2a &=& \frac{k}{\sqrt{E(0)}} + k' + k'' \sqrt{E(0)} \; , \\
\frac{b}{a} &=& \frac{k'}{4} + \frac{k'' \sqrt{E(0)}}{2}  \; .
\end{eqnarray}
As expected, the initial energy 
$E(0)$ appears with both the non-linear terms $k$ and $k''$, 
and hence both $a$ and $b$ depend on the initial amplitude. 
More importantly, the decay slope $a$ depends on 
the damping constants $k$, $k'$, and $k''$, which implies that the
linear part of the amplitude decay itself contains information about 
all the damping terms.

A typical amplitude decay with time is shown in Fig. 2 
for both the open- and short-circuit configurations.
A dominantly linear decay is seen in both cases,
with a deviation from linearity becoming pronounced at large times.
In the open-circuit configuration, the damping is due to friction 
and air resistance. 
When the coil is short-circuited
(through a low resistance ($220\; \Omega$)
so that the emf pulses can be monitored by 
tapping the voltage across the resistor),
intermittent electromagnetic damping is activated due to the 
induced eddy current when the magnet passes through the coil.

Least-square fits with the quadratic form given in Eq. (7) yield 
$a=49\times 10^{-4}$ sec$^{-1}$ and $b=51\times 10^{-7}$ sec$^{-2}$ 
for the open-circuit case,
whereas for the short-circuit case
$a=117\times 10^{-4}$ sec$^{-1}$ and $b=214\times 10^{-7}$ sec$^{-2}$.
The linear decay rate $a$ is significantly larger in the
short-circuit case due to the additional 
contribution to $k$ from electromagnetic damping.
From Eq. (9), we note that the ratio $b/a$ 
depends only on the air-resistance damping coefficients $k'$ and $k''$.
If the electromagnetic damping only modifies the $k$ term,\cite{faraday} 
then this ratio should be indentical for both cases,
provided the initial amplitude is identical.
The near doubling of this ratio in the short-circuit case 
is therefore a clear indication that electromagnetic damping 
also introduces small $E$ and $E\sqrt{E}$ damping terms which 
modify the quadratic coefficient $b$.

\begin{figure}
\vspace*{-70mm}
\hspace*{-38mm}
\psfig{file=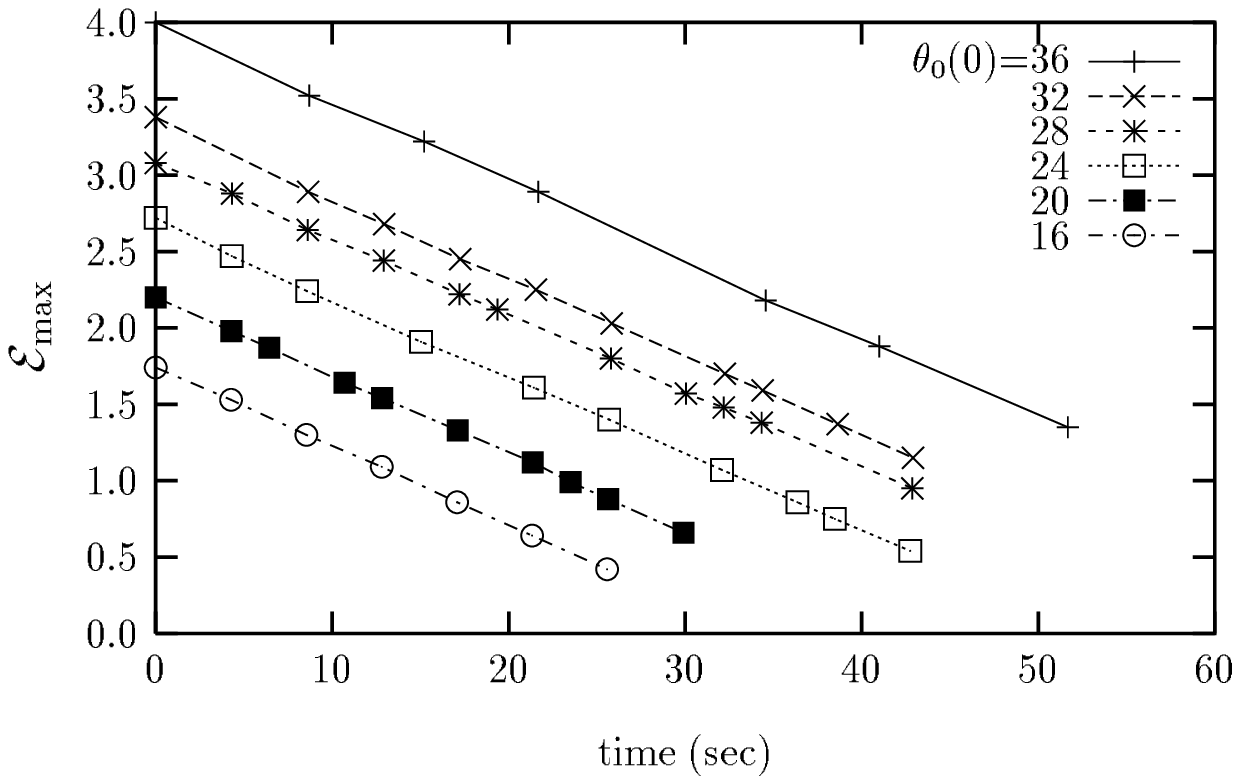,width=135mm,angle=0}
\vspace{-70mm}
\caption{
The decay of the peak emf with time for different initial angular amplitudes
shows no change in the decay slope, 
hence indicating no observable effect of air resistance.}
\end{figure}

\section{Initial-amplitude dependence of the decay slope} 
Equation (9) shows that both $k'$ and $k''$ contribute to the 
coefficient $b$ of the quadratic term in Eq. (7).
Therefore, it is not possible to distinguish between the $E$ and $E\sqrt{E}$ 
damping terms from the initial amplitude decay, 
unless the dependence on the initial energy (amplitude) is taken into account.
The form of Eq. (8) suggests a pedagogically instructive 
experimental procedure for identifying the signature 
of the weak damping terms $k$, $k'$, and $k''$
by studying the initial amplitude dependence. 

If the amplitude decay is written as 
$\theta_0(t)  =  \theta_0 (0) - \tilde{\alpha} t + \tilde{\beta} t^2 $,
and the peak emf ${\cal E}_{\rm max}(t) = K \theta_0 (t)$ as 
\begin{equation}
{\cal E}_{\rm max} (t) = {\cal E}_{\rm max} (0) -  \alpha t + \beta t^2 \; ,
\end{equation}
then from Eqs. (7) and (8),
the slope $\alpha = K \theta_0 (0)\, a $ of the linear part of the decay 
can be written as 
\begin{equation}
\alpha =  \frac{K}{2C} \left [
k + k' \, C \theta_0 (0) +  k'' \, C^2 \theta_0 ^2 (0) \right ] \; ,
\end{equation}
where $\sqrt{E(0)} = C \theta_0 (0)$ defines the constant $C$.
Therefore,  a constant, linear, and quadratic dependence of the slope $\alpha$ 
on the initial amplitude $\theta_0 (0)$ will be signatures of 
the weak damping terms $k$, $k'$ and $k''$, respectively.

Figure 3 shows the amplitude decay 
for different initial amplitudes $\theta_0 (0)$
in the open-circuit configuration.
There is no noticeable change in the slope, 
as confirmed from the plot of the decay slope $\alpha$ vs. $\theta_0 (0)$ 
shown in Fig. 4. 
The slope remains constant at $5.15 \times 10^{-2}$ V/sec,
and this rules out both the damping terms $k'$ and $k''$. 
It thus appears that air resistance and centripetal correction to friction 
do not have any observable contribution to the amplitude decay,
and the friction contribution is given by 
$(K/2C) k = 5.15 \times 10^{-2}$ V/sec.

\begin{figure}
\vspace*{-70mm}
\hspace*{-38mm}
\psfig{file=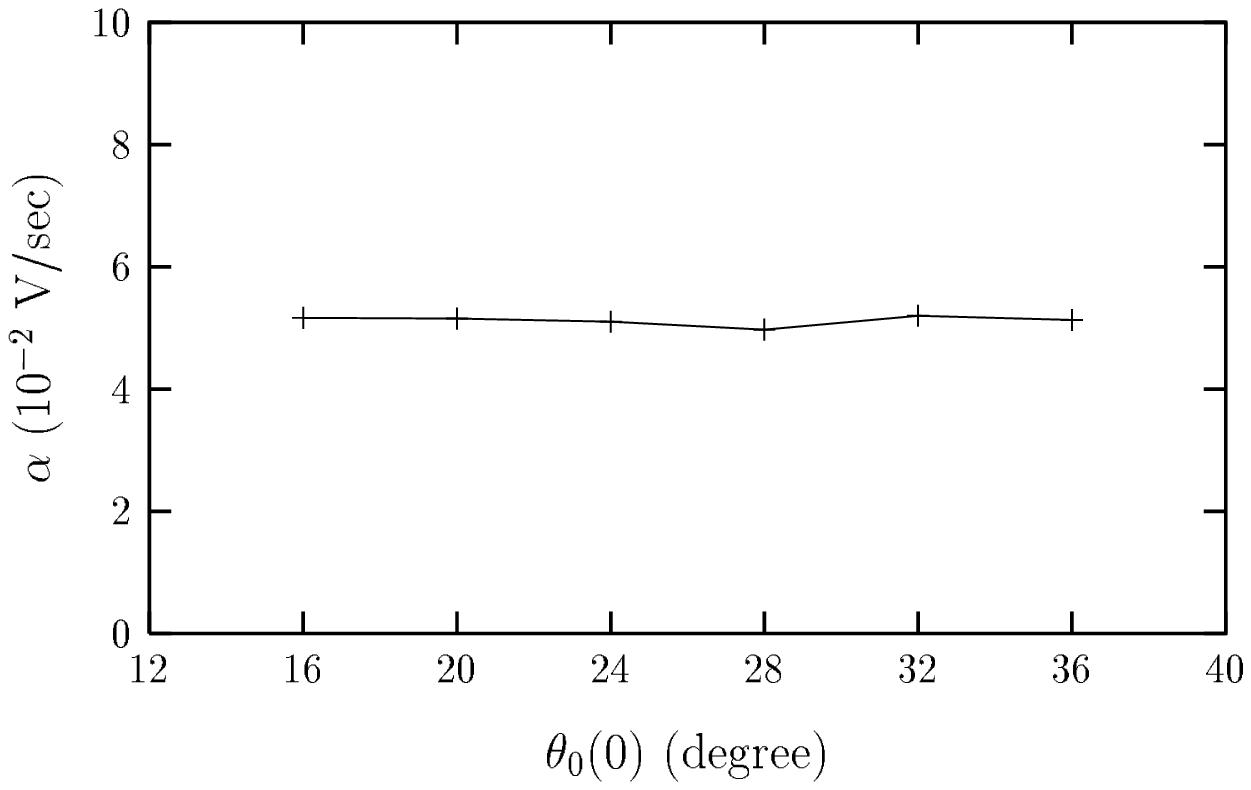,width=135mm,angle=0}
\vspace{-70mm}
\caption{
The decay slope $\alpha$ shows no change with the
initial angular amplitude $\theta_0 (0)$,
showing no observable contribution to the amplitude decay from air resistance.}
\end{figure}

\begin{figure}
\vspace*{-70mm}
\hspace*{-38mm}
\psfig{file=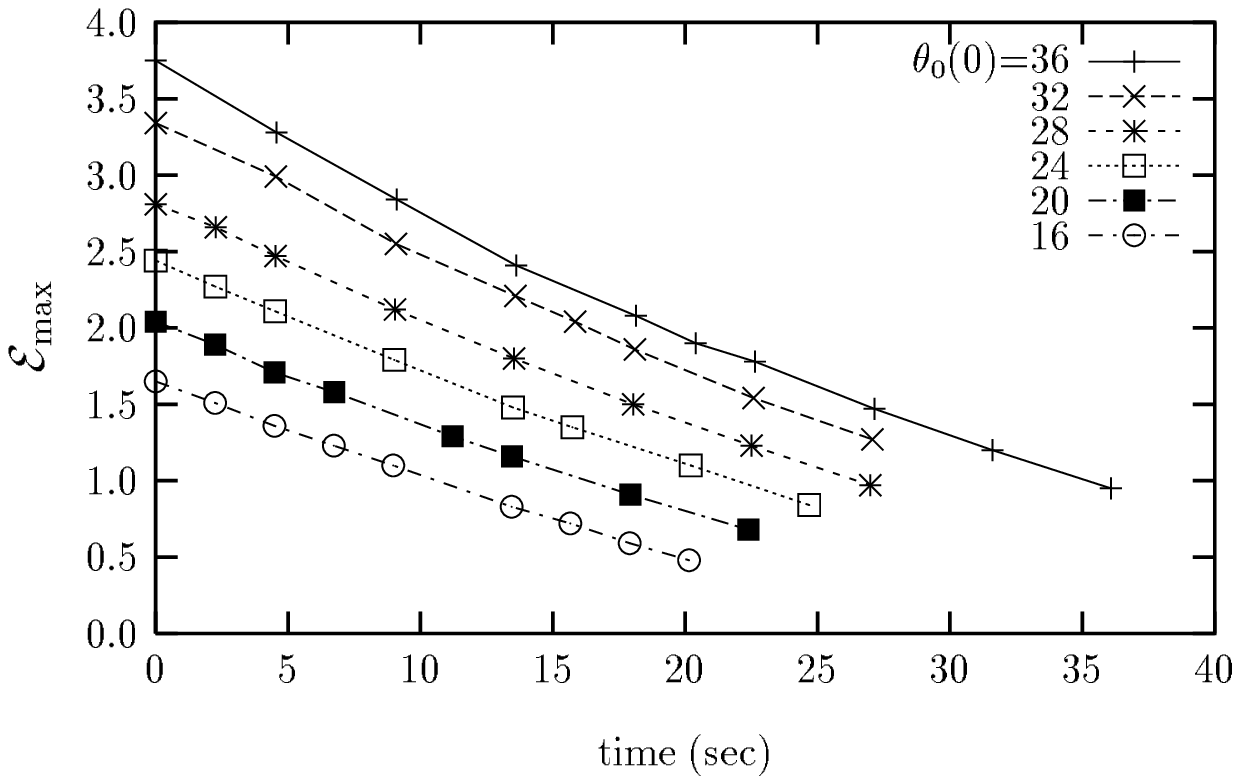,width=135mm,angle=0}
\vspace{-70mm}
\caption{
A variation in the decay slope with the initial angular amplitudes
is clearly observed when a $8\frac{1}{2}'' \times 11 ''$ cardboard sheet is attached 
to the oscillator to provide air resistance.}
\end{figure}

When a $8\frac{1}{2}'' \times 11 ''$ cardboard sheet was attached 
to the oscillator frame to provide air resistance, 
the amplitude decay showed a clear dependence
on the initial amplitude $\theta_0 (0)$, as seen in Fig. 5.
Least-square fit of the peak-emf decay with the quadratic form of Eq. (10)
yields the decay slope $\alpha$, 
which shows a significant increase with $\theta_0 (0)$ (see Fig. 6).
Least-square fit of the decay slope $\alpha$ 
with the quadratic form of Eq. (11)
yields $(K/2C) k' C=3.1 \times 10^{-2}$ V/sec/deg and
$(K/2C)k'' C^2=3.2 \times 10^{-3}$ V/sec/deg$^{2}$
for the linear and quadratic coefficients, respectively.
It is instructive to note that 
the air resistance contribution to the decay slope $\alpha$
due to the cardboard sheet 
is comparable in magnitude to that due to friction at the pivot. 
The presence of a small $k'$ term,
although ideally this term should vanish if the 
air resistance were purely proportional to $v^2$,
has also been noted earlier.\cite{squire}

\begin{figure}
\vspace*{-70mm}
\hspace*{-38mm}
\psfig{file=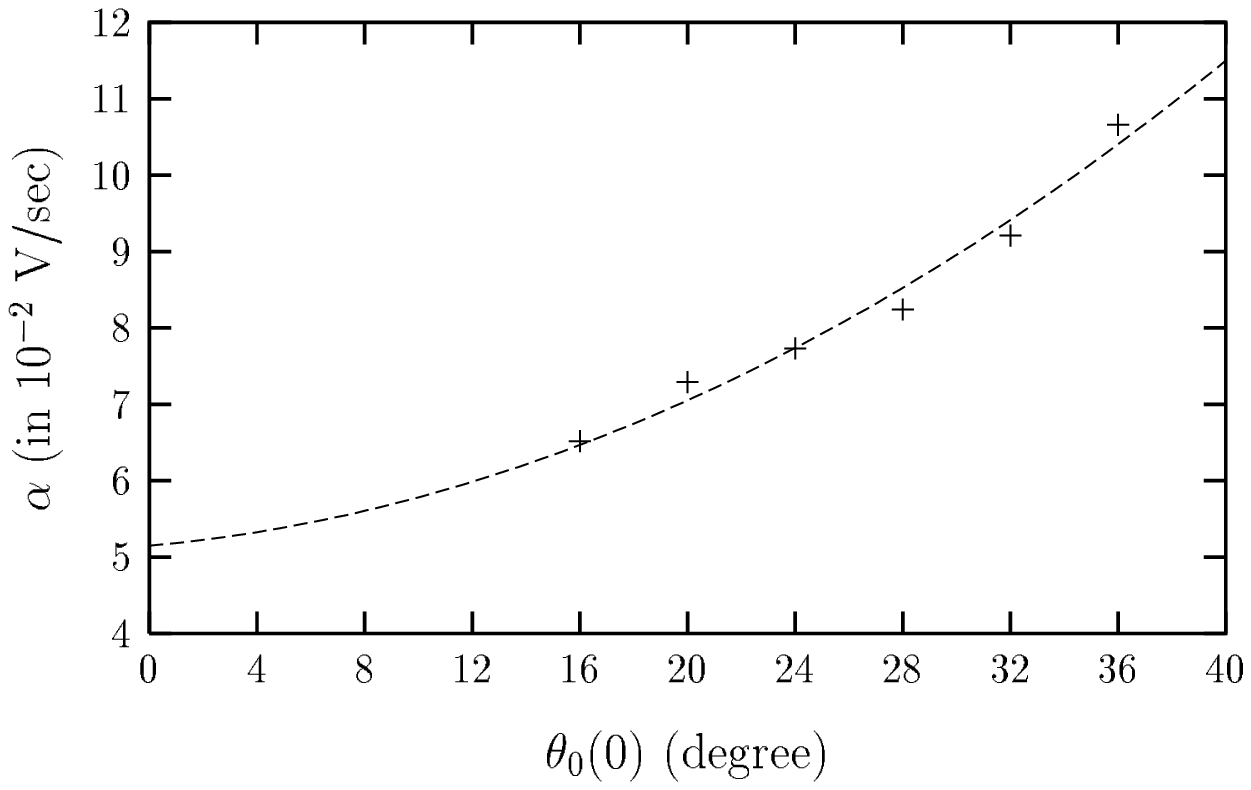,width=135mm,angle=0}
\vspace{-70mm}
\caption{
Increase in the decay slope $\alpha$ with the
initial amplitude $\theta_0 (0)$ is an indicator of the
damping terms $k'$ and $k''$.}
\end{figure}

\section{Summary}
Qualitatively different damping  terms can be distinguished from each other 
by studying the variation of the amplitude decay slope
with the initial amplitude.
A constant, linear, and quadratic dependence of the decay slope  
on the initial amplitude are signatures of the
$E^{1/2}$, $E$, and $E^{3/2}$  damping terms, respectively.
A quantitative determination of the contributions 
of contact friction and air resistance due to an attached vane 
is demonstrated.

\end{document}